\documentstyle{article} 
\input psfig.sty

\begin{document}

{\Large \bf \centerline{Soft X-ray properties of Ultraluminous IRAS Galaxies}}
\vskip 0.5cm
{\it
{\centerline{Th. Boller}}
{\centerline{Max-Planck-Institut f\"ur Extraterrestrische Physik,
Postfach 1603, 85740 Garching}}
}
\vskip 0.5cm

\section{Abstract}
A sample of 323 Ultraluminous IRAS galaxies (ULIRGs) has been correlated
with the ROSAT All-Sky Survey and ROSAT public pointed observations.
22 objects are detected in ROSAT survey observations, and 6 ULIRGs are
detected in addition in ROSAT public pointed observations. The detection is 
based
on a visual inspection of the X-ray contour maps overlaid on optical images
of ULIRGs taken from the Digitized Sky Survey.
Simple power law fits were used to compute the absorption-corrected
fluxes of the  ROSAT detected ULIRGs. The ratio of the soft X-ray flux to the
far-infrared luminosity is used to estimate the contribution
from starburst and AGN emitting processes. These results are compared
with the ISO SWS ULIRG diagnostic diagram.

\section{Class properties of ULIRGs}

\subsection{Specific observational results}

\subsubsection{IRAS 10026+4347}

The ULIRG 
IRAS 10026+4347 can also be classified as
a narrow-line quasar. The  FWHM of the $\rm H\beta$ line is about 2500 $\rm km\ s^{-1}$,
and strong optical Fe II multiplet emission is a prominent feature of
its optical spectrum. The X-ray spectrum exhibits a steep X-ray continuum
slope, with a photon index for a simple power law fit of
$\rm \Gamma = 3.2 \pm 0.5$, typical of narrow-line Seyfert 1 
galaxies and narrow-line quasars. 
The (0.1$-$2.4 keV) luminosity of IRAS 10026+4347, obtained via
a simple power law fit to the data, and corrected for absorption by neutral
hydrogen along the line of sight, is $\rm 1.12 \cdot 10^{45}\ erg \ s^{-1}$.
The ratio of the 
soft X-ray (0.1$-$2.4 keV) to
far-infrared (40$-$120 $\rm \mu$m) flux  is 0.25. In Section 2.3 we argue
that values of this ratio above about 0.003 require a contribution of an
AGN component, in addition to starburst processes.

\subsubsection{Mrk 231}
Mrk 231 (IRAS 12540+5708) is detected both in the ROSAT All-Sky Survey and
in ROSAT public pointed observations. 
The X-ray light curve (cf. Fig. 1), obtained from a 
ROSAT pointed observation, suggests some indication of variability with a
doubling time scale of about 0.4 days. 

\subsection{ROSAT-detected ULIRGs}

22  of the 323 ULIRGs from the IRAS  1.2 Jansky redshift
catalogue (Fisher et al. 1995)
are detected in the 
ROSAT All-Sky Survey (cf. Table 1). 
By inspecting the structure of the X-ray emission in
overlays on optical images taken from the Digitized Sky Survey, it is 
strongly believed,
that the objects in Table 1 are potential identifications of ULIRGs in soft 
X-rays.
Table 2 lists the ULIRGs detected in public ROSAT pointed observations,
in addition to the objects detected in the ROSAT All-Sky Survey.
6 objects are  detected in pointed observations, resulting in a total number of
28 ULIRGs detected with ROSAT. Although the ROSAT energy range does not allow
the probing of  highly obscured regions in ULIRGs, the ROSAT All-Sky Survey allows at
least a statistical approach to the class properties of ULIRGs. 
In Section 2.3 we discuss the ratio of the soft X-ray to far-infrared luminosity
of ROSAT detected ULIRGs, which can be used to estimate the relative fraction
of starburst emitting processes and emission due to accretion onto supermassive
black holes.


\renewcommand{\baselinestretch}{0.8}
\begin{figure}
\centerline{\psfig{figure=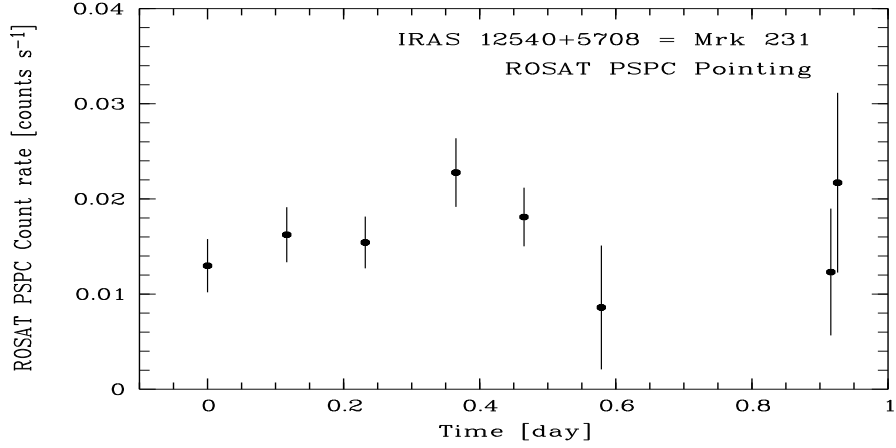,width=12.0cm,height=6.0cm,clip=}}
         \caption{
{\scriptsize
ROSAT PSPC light curve of the ULIRG Mrk 231. The count rate increase suggests
intrinsic variability in the ULIRG Mrk 231.
The estimated upper limit for the size of the emitting region is about 
$\rm 10^{15}\ cm$.
}
}
\end{figure}
\renewcommand{\baselinestretch}{1.0}

\renewcommand{\baselinestretch}{0.8}
\begin{figure}
\mbox{
           \psfig{figure=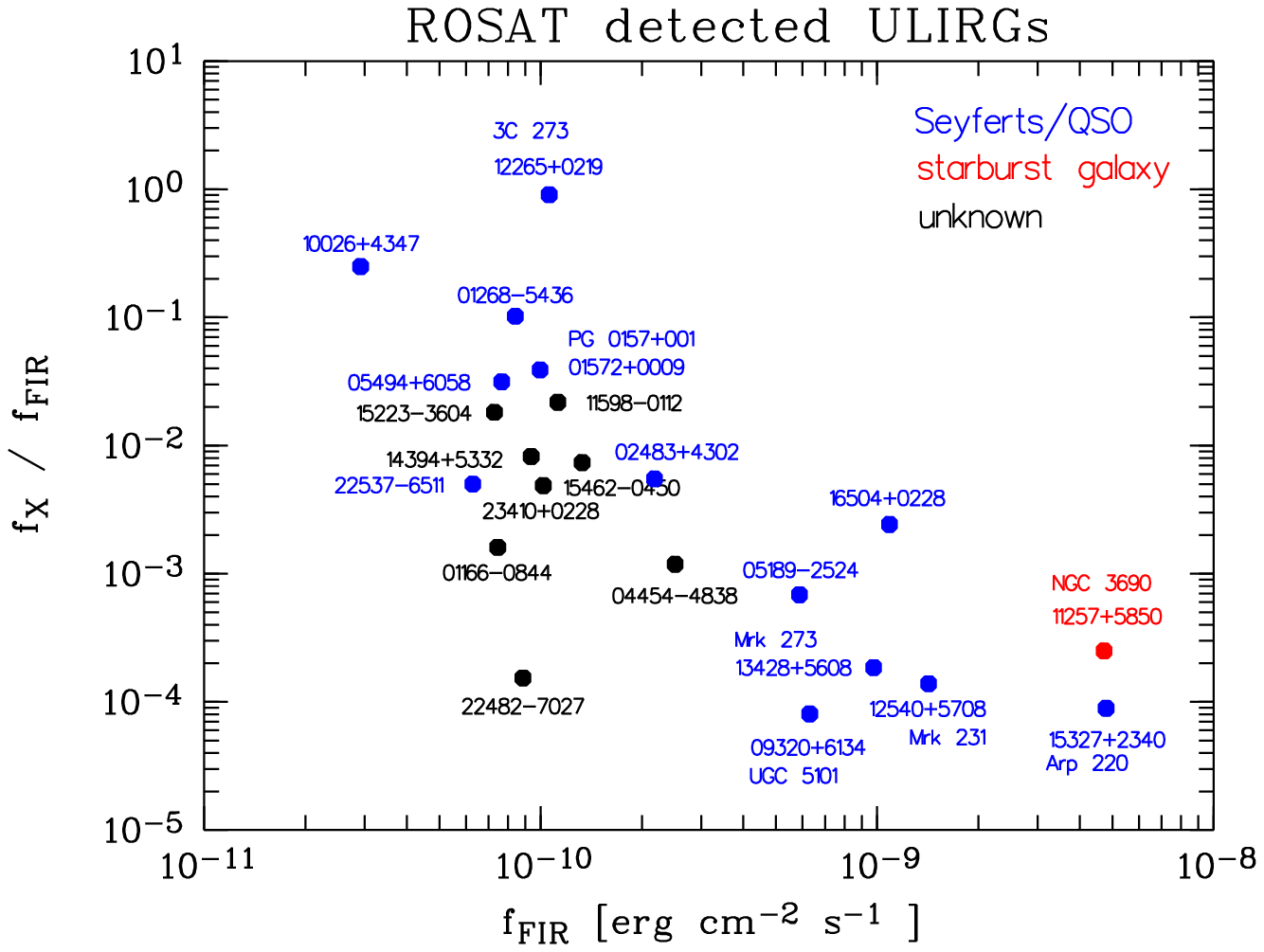,width=8.3cm,height=9.0cm,clip=}
           \psfig{figure=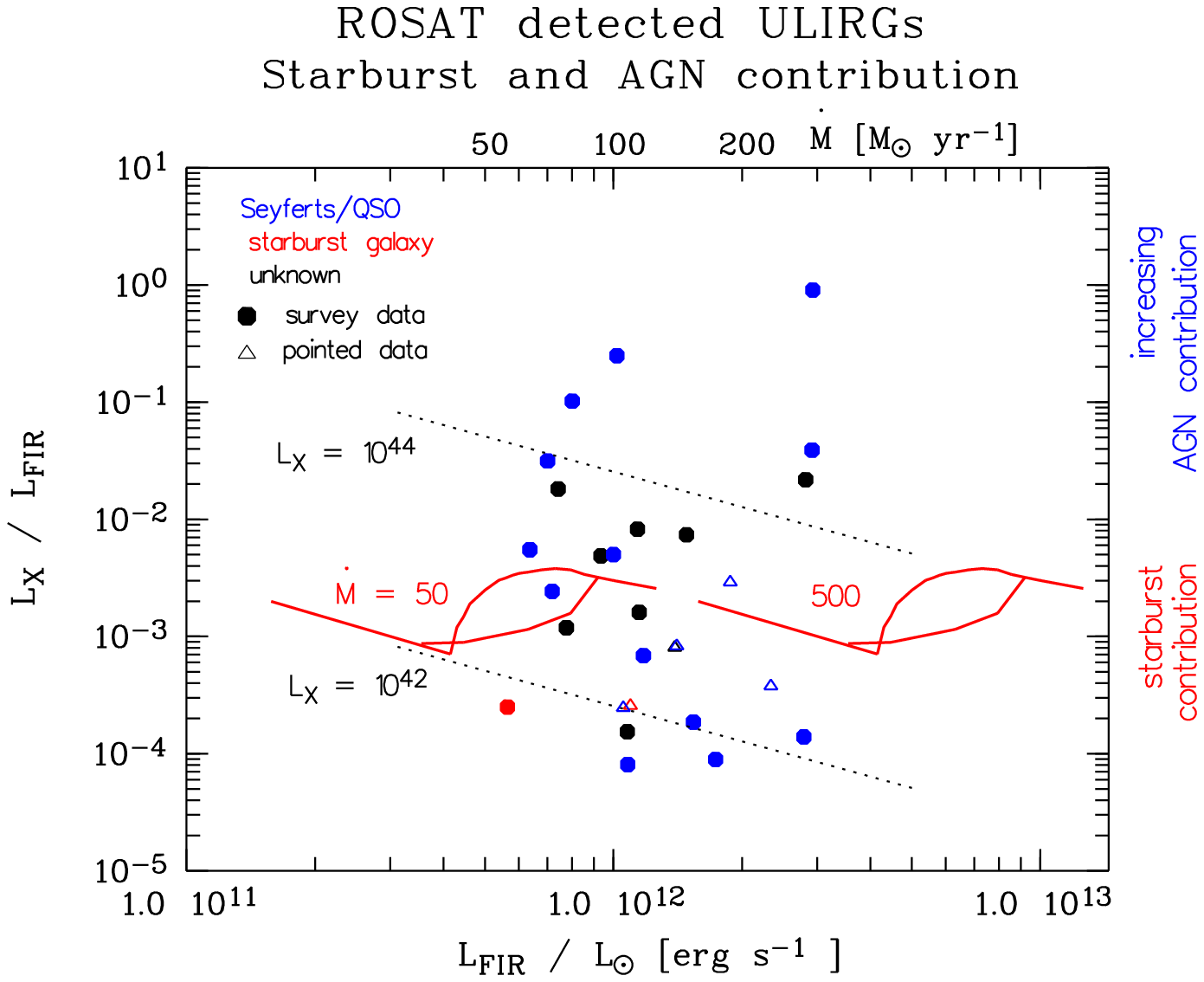,width=8.3cm,height=9.5cm,clip=}}
         \caption{
{\scriptsize
{\bf Left Figure:}
Ratio of the 0.1$-$2.4 keV soft X-ray flux to the far-infrared (40$-$120 $\rm \mu$m)
flux versus the far-infrared flux for ROSAT detected ULIRGs. The $\rm f_X/f_{IR}$
ratio covers 4 orders of magnitude. 
{\bf Right Figure:} Ratio of the soft X-ray luminosity to the far-infrared 
luminosity versus the far-infrared luminosity for ROSAT-detected ULIRGs. The dashed
lines give lines of constant X-ray luminosity. The evolutionary tracks were
obtained from the population syntheses of  star-formation processes with
star-formation rates of 50 and 500 solar masses per year. The mean value of the
soft X-ray to far-infrared luminosity emitted by star-formation processes is about
0.003. Assuming a time-variable star-formation rate, with an increase of the star-formation
rate by a factor of 10 over $\rm 10^8$ years, the ratio between the soft X-ray and the 
far-infrared luminosity varies by about a factor of 3.
This suggests that objects with ratios above about 0.003 are powered by accretion processes
onto a supermassive black hole, in addition to star-formation processes.
}
}
\end{figure}
\renewcommand{\baselinestretch}{1.0}

\begin{table}
\begin{center}
\begin{tabular}{|llll|}
\hline
 (1)  & (2) &  (3) & (4) \\
         IRAS (NED) name   &     log &    log & log \\
            &               $\rm L_{FIR}$ &  $\rm L_X$ & $\rm \frac{L_X}{L_{FIR}}$\\
                            & [erg/s] & [erg/s] & \\
         01166$-$0844 (-)             & 45.65 &42.89& -2.79\\
         01268$-$5436 (-)             & 45.49 &44.53& -0.99\\
         01572+0009  (PG 0157+001)   & 46.05 &44.69& -1.40\\
         02483+4302  (-)             & 45.39 &43.15& -2.25\\
         04454$-$4838 (ESO203-IG1) & 45.48 &42.57& -2.92\\
         05189$-$2524 (-)             & 45.66 &42.51& -3.16\\
         05494+6058 (-)            & 45.43 &43.96& -1.50\\
         11257+5850 (NGC 3690)      & 45.34 &41.74& -3.60\\
         11598$-$0112 (-)            & 46.04 &44.42& -1.66\\
         12265+0219 (3C 273)        & 46.05 &46.06& -0.04\\
         14394+5332 (-)             & 45.64 &43.59& -2.08\\
         15223$-$3604 (-)             & 45.46 &43.75& -1.74\\
         15462$-$0450 (-)             & 45.76 &43.66& -2.13\\
         16504+0228 (NGC 6240)      & 45.44 &42.84& -2.61\\
         22482$-$7027 (-)            & 45.62 &41.84& -3.81\\
         22537$-$6511 (PKS 2253-65)   & 45.59 &43.32& -2.30\\
         23410+0228 (-)            & 45.56 &43.27& -2.31\\
         15327+2340 (Arp 220)       & 45.83 &41.78& -4.04\\
         13428+5608 (MRK 273)       & 45.77 &42.05& -3.73\\
         12540+5708 (Mrk 231)       & 46.03 &42.19& -3.85\\
         09320+6134 (UGC 5101)      & 45.62 &41.54& -4.09\\
         10026+4347 (-)             & 45.59 &45.05& -0.60\\
\hline
\end{tabular}
\end{center}
\caption{
{\scriptsize
ULIRGs detected in the ROSAT All-Sky Survey.
Column 1 lists the
name from the IRAS Point Source Catalogue, and when appropriate, the NED name.
The far-infrared luminosity and the soft X-ray luminosity are listed in columns
2 and 3, respectively. The last column gives the ratio of the soft X-ray
to far-infrared luminosity. This ratio covers about 4 orders of magnitude. 
}
}
\end{table}

\begin{table}
\begin{center}
\begin{tabular}{|llll|}
\hline
 (1)  & (2) &  (3) & (4)    \\
         IRAS (NED) name      & log &  log &   log \\
                          & $\rm L_{FIR}$ &  $\rm L_X$& $\rm \frac{L_X}{L_{FIR}}$\\
                          & [erg/s] & [erg/s] & \\
         07598+6508   &45.86 &43.37 &-2.53\\
         ([HB89]0759+651) & & & \\
         13451+1232 (-)               &45.73 &42.69 &-3.08\\
         14348$-$1447 (-)                   &45.95 &42.56 &-3.42\\
         15033$-$4333 (-)                       &45.73 &42.67 &-3.09\\
         15250+3609  (-)                   &45.61 &42.02 &-3.60\\
         20551$-$4250 (JB40) &45.63 &42.05 &-3.59\\
\hline
\end{tabular}
\end{center}
\caption{
{\scriptsize
ULIRGs detected in public ROSAT pointed observations. 
Objects which are detected in the
ROSAT All-Sky Survey are not included in this table. 
For the description of the columns see Table 1.
}
}
\end{table}

\subsection{The soft X-ray to far-infrared flux ratio}
Since the total X-ray luminosity of a star-forming galaxy is proportional
to its total star-formation rate, one might assume that a high X-ray luminosity might
just reflect a high star-formation rate. A problem with this picture arises when one compares 
the soft X-ray (0.1$-$2.4 keV) flux with the  far-infrared (40$-$120 $\rm \mu$m) flux.
Both quantities are proportional to the star-formation rate, and an increase
in the star-formation rate results in the  first order in a horizontal shift of an
object in Fig. 2.
We (Boller \& Bertoldi 1996) found that in equilibrium, the ratio of the
soft X-ray to far-infrared flux is about 0.003. Considering variable star-formation rates,
the ratio between both quantities varies by about a factor of 3 (see the evolutionary
tracks in the right panel of Fig. 2, where an increase of the star-formation rate
by a factor of 10 for a time scale of $\rm 10^8$ years is assumed). 
The total far-infrared fluxes were computed 
from the 
IRAS 60 and 100 $\rm \mu$m fluxes
following Helou (1985). To compute the soft X-ray fluxes from the PSPC
count rate, a simple power law spectrum of the form $\rm E^{-\alpha}$
was assumed. The fluxes were converted
into luminosities using eq. 7 of Schmidt \& Green (1986). A Hubble constant of 
$\rm H_0 = 50\ km\ s^{-1}\ Mpc^{-1}$ and a cosmological deceleration parameter of
$\rm q_0 = 0.5$ were adopted. In Fig. 2 (left panel) the ratio between the soft X-ray and 
far-infrared
flux  is plotted against the far-infrared flux. The right panel gives
the corresponding distribution for luminosities. The ratio between
the soft X-ray and far-infrared flux ranges over 4 orders of magnitude. 
An additional AGN contribution is necessary to reach the X-ray to 
far-infrared ratio 
for those objects with values above 0.003.

\section{Comparison with ISO SWS results}
\renewcommand{\baselinestretch}{0.8}
\begin{figure}
\centerline{\psfig{figure=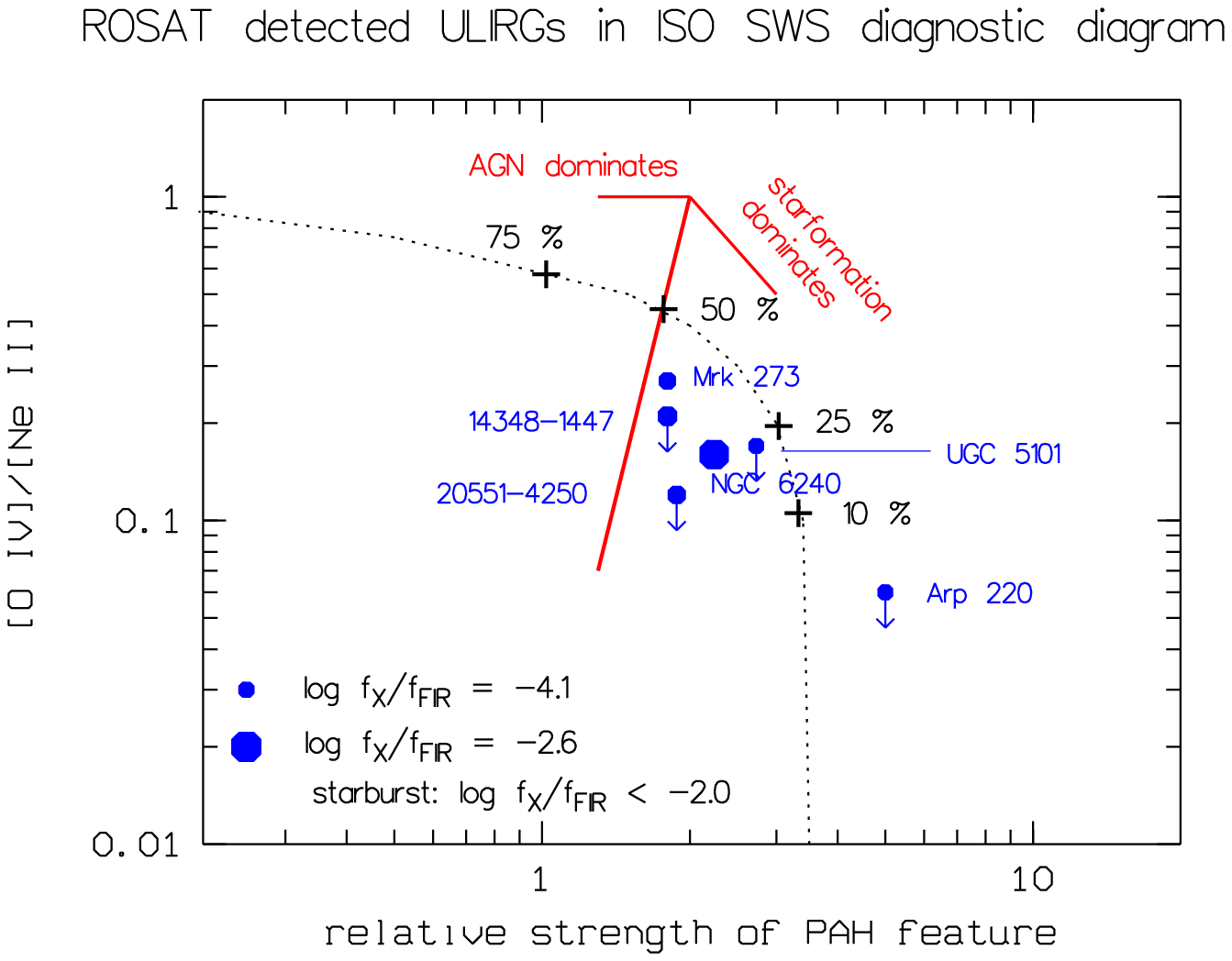,width=11cm,height=9cm,clip=}}
         \caption{
{\scriptsize
ROSAT-detected ULIRGs with known [O IV], Ne II
and PAH feature measurements in the ISO SWS diagnostic diagram. The sizes
of the circles scale with the soft X-ray to far-infrared
flux ratio. All ROSAT-detected ULIRGs in the diagram are predominantly powered
by star-formation processes (see Section 2.3). The ROSAT-detected ULIRGs
are located in that region of the ISO SWS diagram where the AGN contribution
is less than 50 per cent. 
}
}
\end{figure}
\renewcommand{\baselinestretch}{1.0}

In Fig. 3 the ROSAT results are compared with the diagnostic diagram obtained
from ISO SWS measurements to distinguish between starburst and AGN processes
in ULIRGs. The ISO SWS diagnostic diagram shows 
the ratio between the  high- and low-excitation fine structure lines versus the strength of the
PAH  $\rm 7.7\ \mu$m feature (see Lutz, this proceedings). The circles indicate
ROSAT-detected ULIRGs which have measured values in the ISO diagnostic diagram.
The size of the circle scales with increasing soft X-ray to far-infrared
flux ratio. All ROSAT detected-ULIRGs in this diagram are predominantly powered by
star-formation processes, as  the ratio for all objects is below the critical value
of 0.003. This is in agreement with the prediction from the ISO measurements, as all
objects are located in Fig. 3 in the region where the AGN contribution is less
than 50 per cent. 
For ROSAT-detected ULIRGs with ratios of the soft X-ray to far-infrared flux above a
value of 0.003, no ISO SWS [O IV], Ne II or PAH feature measurements 
are available.

\section{Future prospects for the study of ULIRGs}
We have proposed to observe the most interesting
ULIRGs within the guaranteed time program of XMM. We intend
to extend our studies by precisely determining  
the spectral and timing properties of ULIRGs, 
to further disentangle starburst- and AGN-emitting processes in ULIRGs.


{}



\end{document}